# OLIVE OIL BY-PRODUCT AS FUNCTIONAL INGREDIENT IN BAKERY PRODUCTS

**Authors**

Mattia Di Nunzio[a,b], Gianfranco Picone[a], Federica Pasini[b], Elena Chiarello[b], Maria Fiorenza Caboni[a,b], Francesco Capozzi[a,b], Andrea Gianotti[a,b], Alessandra Bordoni[a,b,*]

**Affiliation**

[a]Department of Agri-Food Sciences and Technologies (DISTAL), University of Bologna, Piazza Goidanich 60, 47521 Cesena, Italy

[b]Interdepartmental Centre for Industrial Agri-Food Research (CIRI), University of Bologna, Piazza Goidanich 60, 47521 Cesena, Italy

**\* Corresponding author**

Alessandra Bordoni

Department of Agri-Food Sciences and Technologies (DISTAL), University of Bologna, Piazza Goidanich 60, 47521 Cesena, Italy

e-mail: alessandra.bordoni@unibo.it

Phone: +39 0547338955

---

*Abbreviations:* Ala, alanine; B, blank; BIS, biscuits; CFB, conventional fermented bread; Cho, choline; d6-DSS, 2,2-dimethyl-2-silapentane-d6-5-sulfonic; DOP, defatted olive pomace; Eth, ethanol; Glu, glucose; GPADHEA, 1-β-D-glucopyranosyl acyclodihydroelenolic acid; IFNγ, interferon gamma; IL-10, interleukin 10; IL-12p70, interleukin 12p70; IL-1α, interleukin 1 alpha; IL-1β, interleukin 1 beta; IL-2, interleukin 2; IL-4, interleukin 4; IL-6, interleukin 6; IL-8, interleukin 8; Lac, lactate; LMWF, low molecular weight fraction; myI, myo-inositol; MTT, methylthiazolyldiphenyl-tetrazolium bromide; oPc, o-phosphocholine; q.s., quantum sufficit; SFB, sourdough fermented bread; snG3pc, sn-glycero-3-phosphocholine, TNFα, tumor necrosis factor alpha; US, unsupplemented;

**Abstract**

By-products represent a major disposal problem for the food industry, but they are also promising sources of bioactive compounds. Olive pomace, one of the main by-products of olive oil production, is a potential low-cost, phenol-rich ingredient for the formulation of functional food. In this study, bakery products enriched with defatted olive pomace powder and their conventional counterparts were chemical characterized and *in vitro* digested. The bioaccessible fractions were supplemented to cultured human intestinal cells exposed to an inflammatory stimulus, and the anti-inflammatory effect and metabolome modification were evaluated.

Although in all bakery products the enrichment with olive pomace significantly increased the total phenolic content, this increase was paralleled by an enhanced anti-inflammatory activity only in conventionally fermented bread. Therefore, while confirming olive oil by-products as functional ingredients for bakery food enrichment, our data highlight that changes in chemical composition cannot predict changes in functionality. Functionality should be first evaluated in biological *in vitro* systems, and then confirmed in human intervention studies.

1. INTRODUCTION

Over the past years, researchers and food manufacturers have become increasingly interested in polyphenols. The chief reason for this interest is the recognition of the antioxidant properties of these phytochemicals, their great abundance in the human diet, and their probable role in the prevention of various diseases associated with oxidative stress and inflammation such as cancer, diabetes, cardiovascular and neurodegenerative diseases (Tresserra-Rimbau, Lamuela-Raventos, & Moreno, 2018). Olive oil is an important source of dietary polyphenols, and epidemiological studies have shown an inverse association between its intake and the occurrence of different types of cancer, cardiovascular risk factors, age related processes and chronic inflammatory disorders (Foscolou, Critselis, & Panagiotakos, 2018). These health benefits are mainly attributed to oleuropein, tyrosol and their derivatives that are the major phenolic constituents of olive oil (Karkovic Markovic, Toric, Barbaric, & Jakobusic Brala, 2019).

Olive oil production, an agro-industrial activity of vital economic significance for many Mediterranean countries, is associated with the generation of large quantities of wastes, and olive pomace is one of the main by-product of olive processing (Sinrod et al., 2019). Nowadays, olive pomace is recognized as a potential low-cost starting material rich in bioactive phenolics (Antonia Nunes et al., 2018). In a recent study, we evidenced the positive effect of a polyphenols rich extract from olive pomace on inflammation and cell metabolome in intestinal cell culture (Di Nunzio, Picone, et al., 2018).

The use of olive pomace could represent an innovative and low-cost strategy to formulate healthier and value-added foods, and bakery products are good candidates for enrichment. In fact, they are widely consumed, and a wide variety of bakery product such as bread, biscuits, crackers, breadsticks, and others can be found on supermarket shelves. The nutritional value of bakery products can be improved by different strategies, particularly using by-products as ingredients for the formulation of new functionally enriched food (Boubaker, Omri, Blecker, & Bouzouita, 2016).

The development of innovative bakery products requires knowledge about the influence of processes that could contribute to functionality beside enrichment. As example, sourdough fermentation has been shown to modulate the functional features of leavened baked goods (Gobbetti et al., 2019). As well, the evaluation of the putative functionality of enriched bakery products cannot be simply assessed by evaluating their chemical composition. Modification occurring during digestion and the effect on biological system must be carefully considered.

In this study, bakery products (biscuits and breads) obtained using different flours and fermentation protocols and enriched with defatted olive pomace (DOP) were chemically characterized and *in* *vitro* digested. After digestion, the fraction with a molecular weight compatible to absorption was used for supplementation of intestinal cells. Cultured enterocytes were used as model system since *in vivo* they are in direct contact to the digestion products. Cells underwent exogenous inflammation, and the effect of digested bakery products was investigated measuring cytokines secretion and evaluating the modification of cell metabolome by NMR spectroscopy.

## 2. MATERIAL & METHODS

### 2.1. Chemicals

Dulbecco's modified Eagle's medium (DMEM), penicillin, streptomycin, and Dulbecco's phosphate-buffered saline (DPBS) were purchased from Lonza (Basel, Switzerland). All other chemicals and solvents were of the highest analytical grade from Sigma-Aldrich Co. (St. Louis, MO, USA).

### 2.2. Fermentation and baking processes

Whole einkorn flour (*Triticum monococcum* L. var. Monlis) and whole wheat flour (*Triticum aestivum* L.) were provided by Prometeo (Urbino, Italy). DOP powder was provided by ISANATUR SPAIN S.L. (Puente La Reina, Spain). DOP powder was obtained from the by-product of olive oil extraction by drying and defatting based on the patent WO2013030426 with further developments to increase the process sustainability.

Six different types of bakery products were prepared and tested:
  a) biscuits (BIS) made with whole einkorn flours, with or without 2.5% DOP;
  b) conventional fermented bread (CFB) made with whole wheat flours with or without 4% DOP;
  c) sourdough fermented bread (SFB) made with whole wheat flours with or without 4% DOP.

DOP concentration for enrichment was chosen based on the limit of organoleptic acceptance in consumer preference test (data not shown).

For sourdough preparation, a 10% (fresh weight basis - fwb) mixed-strain containing Lb *plantarum* 98A, Lb. *sanfranciscensis* BB12, Lb. *brevis* 3BHI, and 3% (fwb) of S. *cerevisiae* LBS was added to the dough and fermented at 32 °C for 24 h to obtain a mature sourdough. The microbial load in mature sourdough was approximately $10^9$ colony forming units (CFU)/g for LAB and $10^7$ CFU/g for S. *cerevisiae*. For conventional fermentation, 14% (fwb) of S. *cerevisiae* LBS was added to the final dough for 1.5 h of leavening at 32 °C.

The recipes of the bakery products are reported in Table 1. Biscuits and breads were cooked in oven at 180 °C for 25 minutes and 200 °C for 30 minutes, respectively. After baking, they were cooled at room temperature, cut into small pieces and stored at -18 °C until analysis. Each bakery product was prepared in triplicate on separate days.

*2.3. Polyphenol extraction and determination by HPLC-DAD-MS*

The phenolic fractions were obtained from bakery products by solid-liquid extraction as described by Marzocchi *et al.* (Marzocchi et al., 2017) with slight modifications. Briefly, 3 g of powdered sample were defatted with 30 mL of hexane under stirring for 30 minutes. After removing supernatant, the defatted pellet was extracted by ultrasonic bath at 40 °C for 15 minutes with 30 mL of ethanol/water (4:1 v/v). After centrifugation at 3500 rpm for 15 min, the supernatant was removed, and the extraction was repeated. Supernatants were pooled, evaporated at 35 °C under vacuum and reconstituted with 6 mL of water/methanol (1:1 v/v). The final extracts were filtered with regenerated 0.2 μm cellulose filters (Millipore, Bedford, MA, USA) and stored at -18 °C until analysis.

Polyphenol profile was determined using an Agilent 1200-LC system (Agilent Technologies, Palo Alto, CA, USA) equipped with vacuum degasser, autosampler, binary pump, HP diode-array, UV–VIS detector and mass spectrometer detector as previously described (M. Di Nunzio, M. Toselli, V. Verardo, M. F. Caboni, & A. Bordoni, 2013). Separations were carried out on a reverse phase column Poroshell 120 SB-C18 (3x100 mm, 2.7 μm) from Agilent Technologies (Palo Alto, CA, USA) and the chromatogram was registered at 280 nm. The extracted compounds were identified by analysing UV and MS spectra and quantified by DAD detection. The individual phenolic compounds were quantified by their UV absorbance against external standards using tyrosol, caffeic acid, ferulic acid, chlorogenic acid, oleuropein, verbascoside and rutin at 280 nm for the different class of phenols. Results are expressed as mg/Kg bakery product.

*2.4. Tocol determination by HPLC–FLD*

Lipids were extracted from bakery products as previously reported (Boselli, Velazco, Caboni, & Lercker, 2001). One hundred mg of lipids were dissolved in 1 ml of hexane and filtered through a 0.2 μm nylon filter, then 2.5 μL were injected in a HPLC 1200 series equipped with a fluorimeter detector ($\lambda_{ex}$ = 290 nm, $\lambda_{em}$ = 325 nm) (Agilent Technologies, Palo Alto, CA, USA). Separation of tocopherols was performed by a HILIC Poroshell 120 (3x100 mm, 2.7 μm) from Agilent Technologies (Palo Alto, CA, USA) in isocratic conditions using an n-hexane/ethyl acetate/acetic acid (97.3:1.8:0.9 v/v/v) mobile phase. The flow rate was 0.8 mL/min. Tocopherols were identified by co-elution with the respective standards. The calibration curve used for quantification was constructed with α- tocopherol standard solutions. Results are expressed as mg/Kg bakery product.

*2.5. In vitro digestion*

The digestion process was performed on 50 g of each experimental bakery product or 50 g of water (blank digestion) for 245 min (5 min of oral digestion, 120 min of gastric digestion and 120 min of intestinal digestion) at 37 °C, according to the INFOGEST standardized protocol (Minekus et al., 2014) as described in Valli *et al.* (Valli, Taccari, Di Nunzio, Danesi, & Bordoni, 2018). During *in vitro* digestion, several consecutive enzymatic treatments took place by addition of simulated saliva (containing 75 U/mL α-amylase), simulated gastric juice (containing 2000 U/mL pepsin) at acid pH, and simulated pancreatic juice (containing 10 mM bile and 100 U/mL pancreatin) at neutral pH. After digestion, resulting solutions were centrifuged at 50.000 g for 15 min, and the supernatants filtered with 0.2 μm membranes. To separate compounds which size is small enough to be potentially absorbable through the intestinal mucosa, an aliquot was sequentially ultra-filtered with Amicon Ultra (Millipore, Burlington, MA, USA) at 3 kDa of molecular weight cut-off (low molecular weight fraction, LMWF). Each product was digested three times, and the resulting LMWF were mixed and frozen at -18 °C until use.

*2.6. Caco-2 cell culture and supplementation*

Caco-2 human intestinal cells were maintained at 37 °C, 95% air, 5% $CO_2$ in DMEM supplemented with foetal bovine serum (10% v/v), non-essential amino acids (1% v/v), 100 U/mL penicillin and 100 µg/mL streptomycin. Once a week cells were seeded at 9 x $10^3$ cells/mL into a new 75 $cm^2$ flask, and medium was refreshed every 48h (Antognoni et al., 2017).
For experiments, Caco-2 cells were seeded in 24-well at 1 x $10^5$ cells/well concentration (cytotoxicity and inflammation assays) or in 100 mm Petri dishes at 2.7 x $10^6$ cells/dish (NMR assay) and grown for 21 days. Complete differentiation was assessed by measuring the trans epithelial electric resistance of the cell monolayer using the Millicell ERS apparatus (Millipore, Burlington, MA, USA).

In preliminary experiments, to assess LMWF cytotoxicity in basal conditions (i.e. without any inflammatory stimulus), cells were supplemented for 24h with serum-free DMEM containing different concentration of LMWF (2.5, 5, 10, 20 μl/mL medium). Cytotoxicity was assessed as reported below. Based on results obtained, in further experiments cells were supplemented with the highest LMWF concentration having no cytotoxic effects (5μl of LMWF/mL medium). To avoid interference due to the vehicle, some cells were supplemented with the same amount of LMWF from blank digestion (B), while other cells received no supplementation (unsupplemented - US). Concomitant to LMWF supplementation, inflammation was induced exposing cells to interleukin 1 beta (IL-1β) at 10 ng/mL for 24 h.

*2.7. Cytotoxicity evaluation*

LMWF cytotoxicity was evaluated by the methylthiazolyldiphenyl-tetrazolium bromide (MTT) assay and microscopic examination.
  a) Cell viability was determined by conversion of the MTT salt to its formazan product detected at 560 nm using a Tecan Infinite M200 microplate reader (Tecan, Männedorf, Switzerland) (Di Nunzio, Bordoni, Aureli, Cubadda, & Gianotti, 2018). Cell viability is expressed as optical density (O.D.).
  b) Light microscopy examination of cell morphology and monolayer integrity was performed using an inverted confocal light microscopy model IB (Exacta Optech, Modena, Italy) using 10X, 25X and 40X as magnification.

*2.8. Cytokines secretion*

The level of the pro- and anti-inflammatory cytokines interferon gamma (IFNγ), interleukin 1 alpha (IL-1α), IL-1β, interleukin 2 (IL-2), interleukin 4 (IL-4), interleukin 6 (IL-6), interleukin (IL-8), interleukin 10 (IL-10), interleukin 12 p70 (IL-12p70), and tumour necrosis factor alpha (TNFα) was estimated in the cell media by the multiplex sandwich ELISA Ciraplex (Aushon, Billerica, MA, USA) following the manufacturer's instructions. 96-well plates pre-spotted with protein-specific antibodies were used and luminescent signals were detected by Cirascan™ Imaging System.
The level of IL-8 was also estimated in cell media using AlphaLISA kits (Perkin Elmer Inc., Waltham, MA, USA) following the manufacturer's instructions (Valli et al., 2016). We used 96-microwell plates that were read using an EnSpire™ plate reader (Perkin Elmer Inc., Waltham, MA, USA). Results are expressed as pg/mg protein.

*2.9. Protein content determination*

Cells were washed with cold DPBS, lysed with 500 µL of cold Nonidet P-40 (0.25% v/v in DPBS), incubated on ice with shaking for 30 min and centrifuged at 14.000g for 15 min. Supernatants were collected and protein content was determined by Comassie assay (Di Nunzio, Valli, & Bordoni, 2016), using bovine serum albumin as standard.

*2.10. HR $^1$H-NMR*

Cells were washed with ice-cold DPBS, scraped off, the pellet lysed by sonication and centrifuged at 21.000g for 10 min at 4 °C. Five hundred µl of supernatant were added to 10 µl of a $D_2O$ solution of 100 mM 2.2-dimethyl-2-silapentane-d6-5-sulfonic (d6-DSS) with a final concentration in the NMR tube of 9.09 mM. $^1$H-NMR spectra were recorded at 298 K on a Bruker US+ Avance III spectrometer operating at 600 MHz, equipped with a BBI-z probe and a B-ACS 60 sampler for automation (Bruker BioSpin, Karlsruhe, Germany). The HOD residual signal was suppressed by applying the Carr–Purcell–Meiboom–Gill spin-echo pulse sequence with a pre-saturation sequence. Each spectrum was acquired using 32 K data points over a 7183.908 Hz spectral width (12 ppm) and adding 256 transients. A recycle delay of 5 s and a 90° pulse of 11.190 µs were set up. Acquisition time (2.28 s) and recycle delay were adjusted to be 5 times longer than the longitudinal relaxation time of the protons under investigation, which was not longer than 1.4. The data were Fourier transformed and phase and baseline corrections were automatically performed using TopSpin version 3.0 (Bruker BioSpin, Karlsruhe, Germany). Signals were identified by comparing their chemical shift and multiplicity with Chenomx Profiler software data bank (ver. 8.1, Edmonton, Canada) and data in the literature (Picone et al., 2013).

*2.11. Statistical analysis*

Compositional, cytotoxicity and inflammation data are mean ± standard deviation (SD). In HPLC analysis, statistical differences were determined by the one-way analysis of variance (ANOVA) followed by Tukey's post hoc-test considering p<0.05 as significant. In cell culture experiments, statistical differences were determined by Student's t-test and by ANOVA followed by Tukey's post hoc-test considering p<0.05 as significant. NMR spectra processing and PCA analyses were performed using R computational language (ver. 3.5.3). Each $^1$H NMR spectrum was processed by means of R scripts developed in-house. All other statistical analyses were performed using Prism software ver. 5.0 (GraphPad, San Diego, CA, USA).

## 3. RESULTS

### 3.1. Bakery products characterization

Bakery products nutritional composition is given in Table 2, according to producer's analysis.

The phenolic profile of bakery products is given in Table 3, and a representative UV chromatogram of the SFB4% phenolic compounds is reported in figure 1. Total phenol content was significantly higher in all DOP-enriched products than in corresponding controls, the increase being more evident in SFB than in other products.

Tocol concentration of bakery products is given in table 4, and a representative chromatogram of the BIS2.5% tocol profile is reported in figure 2. Enrichment with DOP caused a significant increase in total tocol content, mainly due to the increase of β-tocopherol concentration, in SFB only.

### 3.2. Effects in Caco-2 cells

#### 3.2.1 Experimental set-up in basal condition

Light microscopic examination revealed a deep change of cellular morphology and a loss of cell monolayer integrity in cells supplemented with 10 μl/mL and 20 μl/mL CFB4% and SFB0% (figure 3). In these conditions, an increase in MTT conversion probably related to an enhanced mitochondrial mass and electron transport system activity due to an increased energy requirements associated with cell homeostasis processes under critical condition (Choi, Roche, & Caquet, 2001; Lee, Yin, Chi, & Wei, 2002) was also observed (figure 4). Since the 5 μl/mL concentration had no effect on cell viability, morphology and monolayer integrity regardless the type of LMWF, it was used for supplementation in further experiments.

To select the most appropriate inflammatory stimulus, Caco-2 cells were supplemented for 24h with two different concentrations of lipopolysaccharide (LPS) (100 ng/ml and 500 ng/ml), IL-1β (10 ng/ml and 50 ng/ml), and TNFα (10 ng/ml and 50 ng/ml) alone or in combination (LPS 100 ng/ml +

IL-1β 10 ng/ml + TNFα 10 ng/ml and LPS 500 ng/ml + IL-1β 50 ng/ml + TNFα 50 ng/ml). The onset of inflammation was assessed by measuring the secretion of IL-6 and IL-8 using the AlphaLISA kit assay (figure 5A and 5B, respectively). IL-1β supplementation at 10 and 50 ng/ml significantly enhanced IL-6 and IL-8 secretion, without any additive effect due to combination with TNFα and LPS. On this basis, 10 ng/ml IL-1β were used as inflammatory stimulus in further experiments.

*3.2.2 Anti-inflammatory effect of bakery products*

In basal condition, secretion of most cytokines was below the detection limit. IL-8 was by far the most represented one, and supplementation with B-LMWF did not affect its concentration (Table 5).
Caco-2 cells were stressed by adding 10 ng/ml IL-1β, and the different LMWF were co-supplemented to evaluate their putative anti-inflammatory effect (Table 5). Inflammation caused a significant increase of IL-6 and IL-8 secretion in both US and B supplemented cells compared to basal counterparts (p<0.001). Comparing inflamed cells, IL-8 secretion was significantly lower in CFB4% and SFB0% cells than in US. In CFB4%, it was also lower than in B supplemented enterocytes. All supplementation except BIS2.5% and CFB0% decreased IL-6 secretion.
A relatively high within-group variability was evidenced evaluating cytokine secretion with multiplex sandwich ELISA Ciraplex. Since intra- and inter-assay reproducibility of multiplex assays is sometimes lower than in singleplex assays (Tighe, Ryder, Todd, & Fairclough, 2015), to better investigate the effect of supplementation we measured the secretion of the most expressed cytokine IL-8 also using a singleplex assay, which almost confirmed previous results (figure 6). In basal condition, supplementation with LMWF from B digestion did not modify IL-8 secretion, which was strongly increased by the exposure to the inflammatory stimulus in both US and B cells compared to corresponding basal values (p<0.001 in both cases). In inflamed condition, all supplementations except CFB0% reduced IL-8 secretion. In CFB4% supplemented cells IL-8 secretion was also lower than in B supplemented ones.

*3.2.3. Effect on the metabolome*

Metabolome analysis was performed on the cell lysate. Before statistics, each $^1$H NMR spectrum was processed by means of scripts developed in-house in R language. Chemical shift referencing was performed by setting the DSS signal to 0.00 ppm. Moreover, the alignment of the spectra was improved where possible using the *i*Coshift tool (Savorani, Tomasi, & Engelsen, 2010) available at http://www.models.life.ku.dk/algorithms/. The following spectral regions were removed prior to data analysis: the regions including only noise (the spectrum edges between 11.00 and 9.00 ppm and between 0.8 and -1.00 ppm) and the NMR signal which is strongly affected by the residual solvent peak (water between 4.50 and 5.00 ppm). After the Fourier transformation and prior to multivariate analysis, data underwent to a pre-statistical improvement. First, spectra were normalized to the unit area to reduce possible dilution effects (Craig, Cloarec, Holmes, Nicholson, & Lindon, 2006). Then, to avoid the effect of peaks misalignments among different spectra due to variations in chemical shift of signals belonging to some titratable acids, a points reduction by the "spectral binning" was performed (Gartland, Beddell, Lindon, & Nicholson, 1991) by subdividing the spectra into 410 bins each integrating 100 data points (0.0183 ppm each). A representative 1H NMR spectrum of SFB4% cell lysate is reported in figure 7.

Principal component analysis (PCA) performed on 410 bins is reported in figure 8. PC1 vs PC2 accounted for 86% of the total variance, which was mainly located along PC1 (74%). In order to determine variables encompassing most of the discriminative information, bins with a loading value greater than 1% of the overall standard deviation of all loading values were selected (Picone et al., 2018) on the most important bins along PC1 and PC2. The main metabolites involved in the discrimination among groups were glucose (Glu), lactate (Lac), sn-glycero-3-phosphocholine (snG3pc), o-phosphocholine (oPc), myo-inositol (myI), choline (Cho), alanine (Ala) and ethanol (Eth).

In basal condition, no differences were detected between US and B supplemented cells. Inflammation caused increased lactate signals in US cells compared to basal counterparts, while no differences were detected in B cells. Supplementation did not cause any modification in the metabolome of inflamed cells, except an increase of Cho and myI and a decrease of snG3pc signals observed in SFB4% cells (Table 6).

## 4. DISCUSSION

Today's society, in which there is a great demand for appropriate nutritional standards, is characterized by rising costs and often decreasing availability of raw materials together with much concern about environmental pollution. Consequently there is a considerable emphasis on the recovery, recycling and upgrading of wastes. This is particularly valid for the food industry in which wastes, effluents, residues and by-products can be recovered and often be upgraded to higher value and useful products (Banerjee et al., 2017). In this study, we focused on the possible exploitation of olive oil by-products as functional ingredient in bakery products.

In conventional products, total phenol content reflected the type of flour and fermentation used for baking. The higher total phenol content observed in CFB0% than SFB0% may be accounted to the degradation of the cell wall structure by microbial enzymes during yeast fermentation. This could have caused the release of the aglycones from their glycoside linked to the fibres (Angelino et al., 2017; Laddomada, Caretto, & Mita, 2015), making them more available for hydroalcoholic extraction (Melini & Acquistucci, 2017). The higher total tocols content in conventional biscuits than breads is justified by the use of extra virgin oil for BIS preparation.

The concentration of the typical olive oil polyphenols was increased in the DOP-enriched food compared to their conventional counterparts. Conversely, in BIS and CFB enrichment with DOP did not increase tocol concentration probably due to defatting, which removes most of vitamin E from olive pomace (Rosello-Soto et al., 2015). Therefore, we assume that the increase of tocol concentration in SFB4% should be attributed to sourdough fermentation, as reported by Gianotti *et al.* (Gianotti et al., 2011).

To better mimic *in vivo* effects, bakery products were *in vitro* digested and the LMWFs were used to supplement Caco-2 cells. To avoid misleading results (Di Nunzio et al., 2017), potential cytotoxicity of LMWF was assessed prior to other experiments. As previously reported (Van De Walle, Hendrickx, Romier, Larondelle, & Schneider, 2010), IL-8 was the main cytokine secreted by Caco-2 cells, and secretion was significantly increased upon the inflammatory stimulus. In inflamed cells, supplementation with blank digesta reduced IL-8 secretion at a similar extent than supplementation with digested bakery products, confirming the anti-inflammatory effects of bile acids (Ward et al., 2017) and indicating that the observed effect was mainly due to the vehicle. Only LMWF from digested CFB4% and SFB0% actively contributed to the overall anti-inflammatory effect. CFB4% was the product having the highest phenol concentration, so its effectiveness is easily explained. Conversely, the anti-inflammatory activity of SBF0% is harder to decipher since its phenol and tocol content was lower than SBF4%, which was ineffective. Bioactive compounds must be released from the food matrix to exert a positive action and their bioaccessibility, i.e. the percentage that is made available for absorption during digestion, is influenced by many different factors including the food matrix and the food processing (Bordoni et al., 2011; Ferranti et al., 2014; Marcolini et al., 2015). We hypothesize that the interactions between matrices (including DOP enrichment) and processing differently modulated bioactive bioaccessibility in conventional and experimental breads. Of note, LMWF from CFB4% and SFB0% and not SFB4% exerted cytotoxic effects when supplemented at highest concentrations, confirming that a higher number of active molecules was released from the matrix.

Using an untargeted approach, we evidenced that inflammation has no effect on cell metabolome in our experimental conditions. As well, in inflamed condition, cell metabolome was not modified by any supplementation except SFB4%, which caused a decrease in snG3pc and an increase in myI and Cho concentration. In a previous study, using a similar approach we evidenced that olive polyphenols supplementation causes a huge dose-related perturbation of Caco-2 cells metabolome (Di Nunzio, Picone, et al., 2018). In that study, the amount of olive polyphenols used for supplementation was in the range 50 – 500 μg/ml medium while in the present study it was 100 – 1000 times lower, and it is evident that polyphenol concentration is a main determinant of metabolome perturbation. Of note, in the present study we describe the effect observed in a more physiological situation i.e. the consumption of an enriched food, which polyphenol concentration was based on organoleptic acceptance of the product.

We speculate that in LMWF of SFB4%, the only one influencing the cell metabolome, the high concentration of tocopherol acted synergistically with polyphenols. In fact, the increase of MyI and Cho, which are important precursors of plasma membrane structured lipids (Tayebati, Marucci, Santinelli, Buccioni, & Amenta, 2015; Thomas, Mills, & Potter, 2016) suggests changes in cell membrane integrity (Ricks et al., 2019; Zeisel, Klatt, & Caudill, 2018). Polyphenols, including olive phenols, incorporate into the lipid bilayer inducing biophysical changes (phospholipid re-packing) and altering the membrane structure (de Granada-Flor, Sousa, Filipe, Santos, & de Almeida, 2019; Verstraeten, Fraga, & Oteiza, 2015). As well, tocopherol influences the phase behaviour of lipid bilayer affecting viscosity characteristics and structural transitions of plasmatic membrane (Belov, Mal'tseva, & Pal'mina, 2011; Wang & Quinn, 1999, 2000).

In conclusion, enrichment of bakery products with a by-product of olive oil production caused a significant increase in the total phenolic content that was paralleled by an increased anti-inflammatory activity only in conventionally fermented bread. Thus, our data confirm that the type of fermentation might influence the functional properties of bread, probably by modifying bioaccessibility of phenolic compounds, as previously reported by Katina *et al.,* (Katina et al., 2012) and Wang *et al*. (Wang, He, & Chen, 2014).

Although formulation of functional food by increasing polyphenol concentration through enrichment with olive oil polyphenols by-products represents an important strategy to improve consumers well-being (Nocella et al., 2018; Serino & Salazar, 2019), results herein reported highlight that the increased concentration of bioactive molecules in the food is not enough to guarantee its functionality, and bioaccessibility must be carefully considered.

Notably, the contribution of the gut microbiota to polyphenols transformation was not considered in our model system, and results obtained *in vitro* do not exactly mirror the *in vivo* effect. Although clinical intervention studies are the gold standard to verify the health effect of foods, *in vitro* studies can give useful preliminary indications and may represent the first step towards the formulation of effective functional food.

**Funding**

This study was supported by the EU Project EcoPROLIVE "Ecofriendly PROcessing System for the full exploitation of the OLIVE health potential in products" (grant agreement no. 635597).

**Author contributions**

M.D.N. performed the experiments on cell culture with E.C. and wrote the draft manuscript; G.P. carried out NMR and multivariate statistical analysis; F.P. performed HPLC analyses; A.B., M.F.C., F.C. and A.G. designed and supervised the study. All Authors critically contributed to the manuscript writing.

**Conflicts of interest**

The authors declare no conflict of interest.

**Table 1.**

Recipes of bakery products. q.s.: quantum sufficit.

| Ingredients | BIS0% | BIS2.5% | CFB0% | CFB4% | SFB0% | SFB4% |
|---|---|---|---|---|---|---|
| Whole wheat flour (g) | 0.0 | 0.0 | 98.6 | 94.6 | 97.8 | 93.9 |
| Whole einkorn flour (g) | 36.3 | 35.4 | 0.0 | 0.0 | 0.0 | 0.0 |
| Extra-virgin olive oil (g) | 8.1 | 7.9 | 0.4 | 0.4 | 0.5 | 0.4 |
| Sodium bicarbonate (g) | 0.4 | 0.4 | 0.0 | 0.0 | 0.0 | 0.0 |
| Potassium hydrogen tartrate (g) | 0.5 | 0.5 | 0.0 | 0.0 | 0.0 | 0.0 |
| Brown sugar | 12.0 | 11.7 | 0.0 | 0.0 | 0.0 | 0.0 |
| Water (ml) | 42.7 | 41.6 | q.s. | q.s. | q.s. | q.s. |
| Dough (g) | 0.0 | 0.0 | 0.0 | 0.0 | 1.7 | 1.7 |
| yeast beer for CF (g) | 0.0 | 0.0 | 1.0 | 0.9 | 0.0 | 0.0 |
| DOP (%) | 0.0 | 2.5 | 0.0 | 4.0 | 0.0 | 4.0 |

**Table 2.**

Nutritional composition of bakery products.

|  | BIS0% | BIS2.5% | CFB0% | CFB4% | SFB0% | SFB4% |
|---|---|---|---|---|---|---|
| Humidity | 4.4 | 3.1 | 5.4 | 6.8 | 5.1 | 5.9 |
| Proteins (g/100g) | 9 | 9.7 | 13 | 12.4 | 12.9 | 12.6 |
| Fats (g/100g) | 14.9 | 15.2 | 3.2 | 3.9 | 3.4 | 3.7 |
| Fibres (g/100g) | 4.5 | 5.5 | 8.1 | 9.0 | 7.1 | 7.6 |
| Ashes (g/100g) | 1.9 | 2.0 | 2.0 | 2.2 | 2.0 | 2.3 |
| Carbohydrates (g/100g) | 64.4 | 65.6 | 68.3 | 65.7 | 69.4 | 68 |
| Energetic value (Kcal/100g) | 439.0 | 447.0 | 370.0 | 366.0 | 374.0 | 371.0 |
| Sodium (mg/Kg) | 1843.0 | 2040.0 | 3530.0 | 3720.0 | 3530.0 | 3210.0 |
| Saturated fats (%) | 13.0 | 13.1 | 16.6 | 15.2 | 16.5 | 15.4 |
| Monounsaturated fats (%) | 76.2 | 75.9 | 58.3 | 66.6 | 59.2 | 64.8 |
| Polyunsaturated fats (%) | 11.0 | 11.1 | 25.1 | 18.2 | 24.3 | 19.8 |

1    **Table 3.**

2    Phenolic profile in bakery products. Data are expressed as mg/Kg bakery product and are mean ± SD of three samples. Statistical analysis was by

3    one-way ANOVA (p <0.001 for each phenol), using Tukey's post-hoc test. Different letters in the same row indicate significant differences (at least

4    p<0.05). GPADHEA: 1-β-D-glucopyranosyl acyclodihydroelenolic acid. RT: retention time expressed in minutes; Der: derivative; Iso: isomers.

| # | RT | Phenolic compounds | Mass data ESI$^-$ [M-H]$^-$ | BIS0% | BIS2.5% | CFB0% | CFB4% | SFB0% | SFB4% |
|---|---|---|---|---|---|---|---|---|---|
| 1 | 0.8 | Quinic acid | 191, 111 | 30.7±0.2$^e$ | 45±2.0$^e$ | 164.2±11.2$^a$ | 128.7±6.0$^b$ | 71.8±3.5$^d$ | 87.9±0.1$^c$ |
| 2 | 1.1 | Gallic acid der. | 305 | 14.9±0.0$^{de}$ | 20.3±0.3$^{cd}$ | 78.5±7$^a$ | 67.2±2.9$^b$ | 11.6±1.1$^e$ | 26.0±0.1$^c$ |
| 3 | 1.3 | Cumaroyl quinic acid | 337 | 14.4±0.3$^c$ | 15.3±0.2$^b$ | 26.5±0.6$^a$ | 15.0±1.0$^c$ | 12.0±0.8$^d$ | 8.8±0.7$^e$ |
| 4 | 1.5 | Feruloyl quinic acid | 367 | 15.1±0.0$^e$ | 36.8±0.8$^d$ | 178.1±9.3$^a$ | 147.3±1.8$^b$ | 94.9±1.3$^c$ | 102.4±4.3$^c$ |
| 5 | 2.5 | Hydroxytyrosol /GPADHEA | 407 | 13.3±1.2$^e$ | 73.6±3.2$^c$ | 38.1±1.9$^b$ | 206.8±4.6$^a$ | 18.7±0.4$^e$ | 162.1±3.2$^b$ |
| 6 | 3.1 | Caffeoyl quinic der. | 353 | 2.5±0.0$^c$ | 5.9±0.0$^c$ | 80.5±11.4$^a$ | 70.4±0.3$^a$ | 5.3±0.5$^c$ | 23.0±0.4$^b$ |
| 7 | 3.4 | Caffeoyl quinic der. | 353 | 116.0±1.0$^a$ | 100.5±4.5$^b$ | 0.0±0.0$^c$ | 0.0±0.0$^c$ | 0.0±0.0$^c$ | 0.0±0.0$^c$ |
| 8 | 4.5 | Caffeoyl quinic der. | 353 | 0.0±0.0$^c$ | 0.0±0.0$^c$ | 7.0±0.7$^b$ | 7.1±0.3$^b$ | 6.8±0.1$^b$ | 12.2±0.2$^a$ |
| 9 | 5.9 | Flavonoid glucoside | 481, 449 | 0.0±0.0$^d$ | 0.0±0.0$^d$ | 12.8±0.9$^a$ | 10.2±1.1$^b$ | 7.3±0.3$^c$ | 11.9±1.5$^a$ |
| 10 | 6.2 | Pinoresinol der. | 357 | 0.0±0.0$^d$ | 0.0±0.0$^d$ | 66.1±1.3$^b$ | 76.6±5.4$^a$ | 37.2±1.5$^c$ | 61.0±3.3$^b$ |
| 11 | 7.2 | Ferulic acid der. | 389 | 0.0±0.0$^d$ | 0.0±0.0$^d$ | 2.9±0.2$^a$ | 2.2±0.2$^b$ | 0.7±0.0$^c$ | 0.6±0.1$^c$ |
| 12 | 8.3 | Hydroxyverbascoside iso. | 623 | 8.4±0.0$^a$ | 7.3±0.2$^b$ | 7.5±0.1$^b$ | 7.1±0.6$^b$ | 3.4±0.2$^c$ | 7.1±0.3$^b$ |
| 13 | 8.8 | Ferulic acid | 193 | 2.5±0.0$^b$ | 3.3±0.1$^a$ | 0.0±0.0$^c$ | 0.0±0.0$^c$ | 0.0±0.0$^c$ | 0.0±0.0$^c$ |
| 14 | 9.3 | Oleuropein der. | 391 | 0.0±0.0$^d$ | 0.0±0.0$^d$ | 88.9±9.4$^b$ | 122.9±12.4$^a$ | 25.8±0.0$^c$ | 104.5±2.7$^b$ |
| 15 | 9.7 | Ferulic acid der. | 371, 193 | 0.0±0.0$^c$ | 0.0±0.0$^c$ | 1.0±0.0$^b$ | 1.3±0.2$^a$ | 1.3±0.0$^a$ | 1.3±0.2$^a$ |
| 16 | 10.1 | Luteolin-7-glucoside | 447 | 0.0±0.0$^c$ | 0.0±0.0$^c$ | 0.0±0.0$^c$ | 4.3±0.0$^a$ | 0.0±0.0$^c$ | 2.2±0.0$^b$ |
| 17 | 11.2 | Verbascoside | 623 | 8.2±0.5$^a$ | 7.1±0.3$^b$ | 4.2±0.1$^c$ | 7.6±0.5$^{ab}$ | 2.6±0.0$^d$ | 4.5±0.2$^c$ |
| 18 | 13.2 | Diferulic acid | 385, 193 | 0.0±0.0$^d$ | 0.9±0.1$^c$ | 0.0±0.0$^d$ | 2.1±0.1$^a$ | 0.0±0.0$^d$ | 1.5±0.2$^b$ |
|  |  | Total |  | 226.0±2.9$^e$ | 316.0±2.6$^d$ | 756.1±53.0$^b$ | 876.7±33.5$^a$ | 299.6±5.8$^d$ | 617.2±6.9$^c$ |

Table 4.

Tocol profile in bakery products. Data are expressed as mg/Kg bakery product and are mean ± SD of three samples. Statistical analysis was by one-way ANOVA (p <0.001 for each tocol), using Tukey's post-hoc test. Different letters in the same row indicate significant differences (at least p <0.05). RT: retention time expressed in minutes.

| # | RT | Tocols | BIS0% | BIS2.5% | CFB0% | CFB4% | SFB0% | SFB4% |
|---|---|---|---|---|---|---|---|---|
| 1 | 1.5 | α-tocopherol | 27.5±1.1$^a$ | 26.9±0.6$^a$ | 2.1±0.1$^b$ | 2.9±0.5$^b$ | 1.5±0.0$^b$ | 2.8±0.2$^b$ |
| 2 | 1.8 | α-tocotrienol | 4.7±0.3$^a$ | 4.1±0.1$^b$ | 0.6±0.0$^c$ | 0.4±0.0$^c$ | 0.5±0.0$^c$ | 0.4±0.0$^c$ |
| 3 | 2.0 | β-tocopherol | 19.7±1.1$^a$ | 21.5±0.8$^a$ | 4.2±0.2$^d$ | 4.7±0.3$^d$ | 8.8±0.2$^c$ | 15.2±1.0$^b$ |
| 4 | 2.5 | β-tocotrienol | 7.2±0.7$^a$ | 6.4±0.2$^a$ | 3.4±0.1$^b$ | 3.2±0.2$^b$ | 2.9±0.1$^b$ | 3.3±0.2$^b$ |
|   |   | Total | 59.1±3.2$^a$ | 59±1.3$^a$ | 10.3±0.2$^d$ | 11.2±1.1$^{cd}$ | 13.6±0.3$^{cd}$ | 21.7±1.5$^b$ |

**Table 5.**

Cytokine secretion after 24h of supplementation. Data are expressed as pg/mg of protein and are mean ± SD of four samples from two independent experiments. Statistical analysis was performed by the Student's t-test to compare US and B cells in basal condition († $p<0.05$), and US and B inflamed cells with the corresponding basal counterparts (IL-6 and IL-8: $p<0.001$). Inflamed supplemented cells were compared by one-way ANOVA (IL-6 and IL-8: $p<0.001$) using Tukey's post-hoc test. Different letters in the same row indicate significant differences (at least $p <0.05$). n.d. = not detectable. The limit of detection was 4.75 fg/ml, 0.177 pg/ml, 40.5 fg/ml, 9.72 fg/ml, 59 fg/ml, 0.25 pg/ml, 0.74 pg/ml for IFNγ, IL-1α, IL-2, IL-4, IL-10, IL-12p70, and TNFα, respectively.

|          | US        | B                   | US                      | B                        | BIS0%                    | BIS2.5%                 | CFB0%                   | CFB4%                   | SFB0%                   | SFB4%                    |
|----------|-----------|---------------------|-------------------------|--------------------------|--------------------------|-------------------------|-------------------------|-------------------------|-------------------------|--------------------------|
|          |           |                     | IL-1β (10ng/mL)         |                          |                          |                         |                         |                         |                         |                          |
| IFNγ     | n.d.      | n.d.                | n.d.$^a$                | n.d.$^a$                 | n.d.$^a$                 | n.d.$^a$                | n.d.$^a$                | 0.00±0.01$^a$           | n.d.$^a$                | 0.01±0.01$^a$            |
| IL-1α    | n.d.      | 0.34±0.39           | n.d.$^a$                | n.d.$^a$                 | n.d.$^a$                 | 0.27±0.44$^a$           | n.d.$^a$                | 0.37±0.52$^a$           | n.d.$^a$                | 0.3±0.29$^a$             |
| IL-2     | n.d.      | 0.01±0.01           | n.d.$^a$                | n.d.$^a$                 | n.d.$^a$                 | 0.02±0.02$^a$           | n.d.$^a$                | 0.01±0.02$^a$           | n.d.$^a$                | 0.03±0.03$^a$            |
| IL-4     | n.d.      | n.d.                | n.d.$^a$                | n.d.$^a$                 | n.d.$^a$                 | 0.01±0.01$^a$           | n.d.$^a$                | n.d.$^a$                | n.d.$^a$                | 0.01±0.02$^a$            |
| IL-6     | 0.05±0.05 | 0.15±0.05$^\dagger$ | 0.73±0.32$^{ab}$        | 0.56±0.17$^{bcd}$        | 0.61±0.24$^{bcd}$        | 0.82±0.13$^{ab}$        | 1.11±0.26$^a$           | 0.14±0.1$^d$            | 0.2±0.22$^{cd}$         | 0.55±0.12$^{bcd}$        |
| IL-8     | 22.9±2.3  | 21±5.8              | 175.3±44.4$^a$          | 137.1±16.7$^{ab}$        | 139.9±22.9$^{ab}$        | 127.2±4.3$^{abc}$       | 173.6±37.1$^a$          | 62.9±20.7$^c$           | 68.5±52$^{bc}$          | 127.4±30.4$^{abc}$       |
| IL-10    | n.d..     | 0.01±0.03           | n.d.$^a$                | 0.01±0.02$^a$            | n.d.$^a$                 | 0.06±0.09$^a$           | n.d.$^a$                | 0.02±0.03$^a$           | n.d.$^a$                | 0.03±0.42$^a$            |
| IL-12p70 | n.d.      | n.d.                | n.d.$^a$                | n.d.$^a$                 | n.d.$^a$                 | n.d.$^a$                | n.d.$^a$                | n.d.$^a$                | n.d.$^a$                | 0.25±0.42$^a$            |
| TNFα     | n.d.      | n.d.                | n.d.$^a$                | n.d.$^a$                 | n.d.$^a$                 | n.d.$^a$                | n.d.$^a$                | n.d.$^a$                | n.d.$^a$                | n.d.$^a$                 |

**Table 6.**

Integrals of bins from PC loadings. In each condition, data are mean ± SD of three samples coming from independent experiments. Statistical analysis was performed by the Student's t-test to compare US and B cells in basal condition (n.s.), and US and B inflamed cells with the corresponding basal condition (‡ $p<0.05$). Inflamed supplemented cells were compared by one-way ANOVA (snG3pc: $p<0.01$; myI and Cho: $p<0.001$) using Tukey's post-hoc test. Different letters in the same row indicate significant differences (at least $p<0.05$).

|        | US        | B         | US                    | B                   | BIS0%              | BIS2.5%            | CFB0%              | CFB4%               | SFB0%              | SFB4%              |
|--------|-----------|-----------|-----------------------|---------------------|--------------------|--------------------|--------------------|---------------------|--------------------|--------------------|
|        |           |           |                       |                     |                    | IL-1β (10 ng/mL)   |                    |                     |                    |                    |
| Glu    | 29.5±1.8  | 30.4±5.9  | 28.1±0.85$^a$         | 28.1±3.6$^a$        | 31.8±4.2$^a$       | 26.9±5.4$^a$       | 31.2±2.2$^a$       | 35.3±0.85$^a$       | 32.3±3.4$^a$       | 28.3±0.4$^a$       |
| Lac    | 57.6±3.2  | 58.6±4.1  | 63.1±0.9$^{‡a}$       | 58.4±1.6$^a$        | 57.1±9.5$^a$       | 63.7±2.1$^a$       | 59.5±2.3$^a$       | 62.1±3.1$^a$        | 63.7±3.8$^a$       | 54.5±12.2$^a$      |
| snG3pc | 47.8±4.74 | 45.1±6.6  | 47.1±2.5$^a$          | 44.5±6.4$^a$        | 43.2±1.6$^{ab}$    | 41.5±3.1$^{ab}$    | 46.5±2.5$^a$       | 40.1±4.9$^{ab}$     | 41.3±5.6$^{ab}$    | 34.4±3.5$^b$       |
| oPc    | 47.3±4.1  | 47.7±5.4  | 45.3±1.2$^a$          | 45.4±1.9$^a$        | 45.5±1.9$^a$       | 43.8±4.5$^a$       | 50.1±3.2$^a$       | 43.6±3.5$^a$        | 44.1±5.9$^a$       | 41.8±0.8$^a$       |
| myI    | 10.7±0.7  | 12.1±1.7  | 9.7±1.7$^b$           | 13.1±2.2$^b$        | 11.6±1.6$^b$       | 11.9±2.4$^b$       | 12.2±1.1$^b$       | 14.1±3.5$^{ab}$     | 11.8±1.7$^b$       | 19.3±0.8$^a$       |
| Cho    | 72.2±3.5  | 88.9±18.5 | 64.1±10$^b$           | 94.5±24.9$^b$       | 83.7±9.2$^b$       | 89.6±13$^b$        | 94.1±7$^b$         | 115.6±35.5$^{ab}$   | 90.9±11.7$^b$      | 166.1±17$^a$       |
| Ala    | 19.2±3.8  | 19.8±5.1  | 22.8±3.1$^a$          | 21.9±1.4$^a$        | 21±7.5$^a$         | 25.5±3.2$^a$       | 21±1.9$^a$         | 23±2.4$^a$          | 24.8±4.3$^a$       | 25.3±2.4$^a$       |
| Eth    | 15.3±1.7  | 16.7±3.1  | 18.8±7.1$^a$          | 15.5±5$^a$          | 15.7±4.8$^a$       | 15.5±1.8$^a$       | 16.6±4.3$^a$       | 13.9±2.5$^a$        | 15.5±2.9$^a$       | 14.6±3.3$^a$       |

**Figure Captions**

**Figure 1.** Chromatogram of the phenolic profile of SFB4%.
Peaks:1) Quinic acid; 2) Gallic acid der.; 3) Cumaroyl quinic acid; 4) Feruloyl quinic acid; 5) Hydroxytyrosol/GPADHEA; 6) Caffeoyl quinic der.; 8) Caffeoyl quinic der.; 9) Flavonoid glucoside; 10) Pinoresinol der.; 11) Ferulic acid der.; 12) Hydroxyverbascoside iso.; 14) Ferulic acid; 15) Ferulic acid der.; 16) Luteolin-7-glucoside; 17) Verbascoside; 18) Diferulic acid;

**Figure 2.** Chromatogram of the tocol compounds of BIS2.5%.
Peaks:1) α-Tocopherol; 2) α-Tocotrienol; 3) β-Tocopherol; 4) β-Tocotrienol.

**Figure 3.** Microscopic observation of cell morphology and layer integrity after 24h of supplementation in basal condition.
A and B: CFB4% and SFB0% supplemented at 10 μl/mL concentration; C and D: CFB4% and SFB0% supplemented at 20 μl/mL concentration. Each picture is representative of six samples from two independent experiments and is taken at 25X of magnification.

**Figure 4.** Cell viability by MTT assay after 24h of supplementation in basal condition.
Cell viability is expressed as optical density (O.D.). Data are mean ± SD of six samples obtained from two independent experiments. Statistical analysis was by the one-way ANOVA (p<0.001) using Dunnett's as post-test to compare supplemented cells to US ones (# p<0.05; ° p<0.01; * p<0.001).

**Figure 5.** IL-6 (A) and IL-8 (B) secretion with different inflammatory stimuli.
Data are expressed as pg/mg of protein and are mean ± SD of four samples from two independent experiments Statistical analysis was by one-way ANOVA (p <0.001 for each IL), using Tukey's post-hoc test. Different letters in the same row indicate significant differences (at least p <0.05).

**Figure 6.** IL-8 secretion after 24h of supplementation.
Data are expressed as pg/mg of protein and are mean ± SD of four samples from two independent experiments. Statistical analysis was performed by the Student's t-test to compare US and B cells in basal condition (n.s.), and US and B inflamed cells with the corresponding basal counterparts (p<0.001 in both cases). Inflamed supplemented cells were compared by one-way ANOVA (p<0.001) using Tukey's post-hoc test. Different letters indicate significant differences (at least p <0.05).

**Figure 7.** Representative $^1$H NMR spectrum in the upfield and midfield region (-0.5:4.60) (panel A) and downfield region (5.00:9.00) (panel B) of SFB4% cell lysate after 24h of supplementation acquired with 600.13 MHz spectrometer at pH 7.33.
1) Isoleucine (t: 0.929, d: 1.000); 2) Leucine (t: 0.946); 3) Valine (d: 0.979, d: 1.032); 4) Ethanol (t: 1.175); 5) Lactate (d: 1.319, q: 4.106); 6) Alanine (d: 1.469); 7) Lysine (m: 1.716, m: 1.908, t: 3.023); 8) Glutamate (m: 2.042, m: 2.121, m: 2.360); 9) Glutathione (q: 2.167, m: 2.565, m:2.977); 10) Creatine (s: 3.028); 11) Ethanolamine (t: 3.138); 12) Choline (s: 3.193), 13) o-Phosphocholine (s: 3.209, m: 3.582, m: 4.156); 14) sn-Glycero-3-phosphocholine (s: 3.218, m: 3.679, m: 4.316), 15) Taurine (t: 3.249, t: 3.416); 16) Glucose (m: 3.395:3.527, m: 3.705:3.894); 16a) β-Glucose (d: 4.640); 16b) α-Glucose (d:5.227); 17) Glucose-6-phosphate (t: 3.274, d: 4.640, d: 5.227); 18) myo-Inositol (t: 3.272, dd:3.528, t:3.615, t: 4.056); 19) Glycine (s: 3.553); 20) O-Phopshoethanoalmine (m: 3.971); 21) 1,3-Dihydroxyacetone (s: 4.413); 22) Uracil (d;5.796, d: 7.528); 23) Uridine (m: 5.905, d: 7.861); 24) UMP (m: 5.975, d:8.101); 25) AMP (d: 6.130, s: 8.260, s: 5.596); 26) Fumarate (s: 6.510); 27) Tyrosine (d: 6.890, d: 7.185); 28) Histamine (s: 7.059, s: 7.794); 29) Phenylalanine (d: 7.322, t: 7.369, t: 7.421); 31) Hypoxanthine (s: 8.181, s: 8.201); 32) Formate (s: 8.445); 33) Nicotinurate (q: 7.590, m: 8.243, dd: 8.705, d:8.931). s: singlet. dd: doublet of doublets. d: doublet. t: triplet. m: multiplet.

**Figure 8.** PCA scores plot of the $^1$H NMR spectrum for all samples (panel A) and representation of the mean values for each group (panel B).

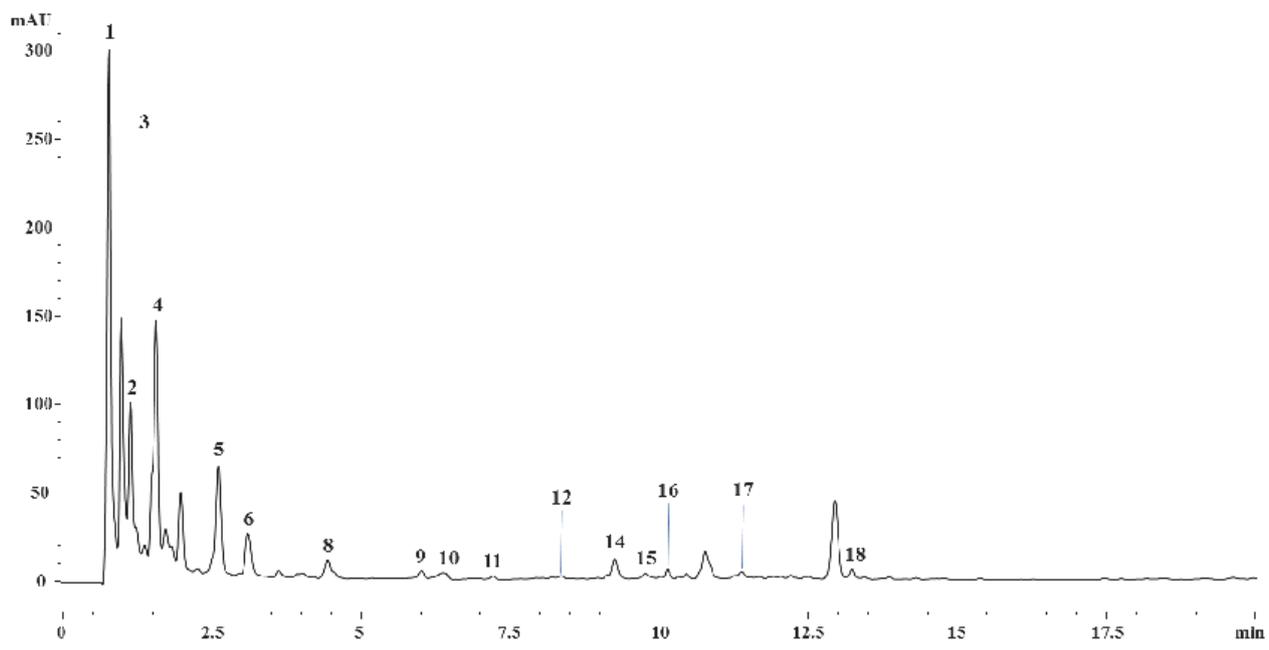

Figure 1.

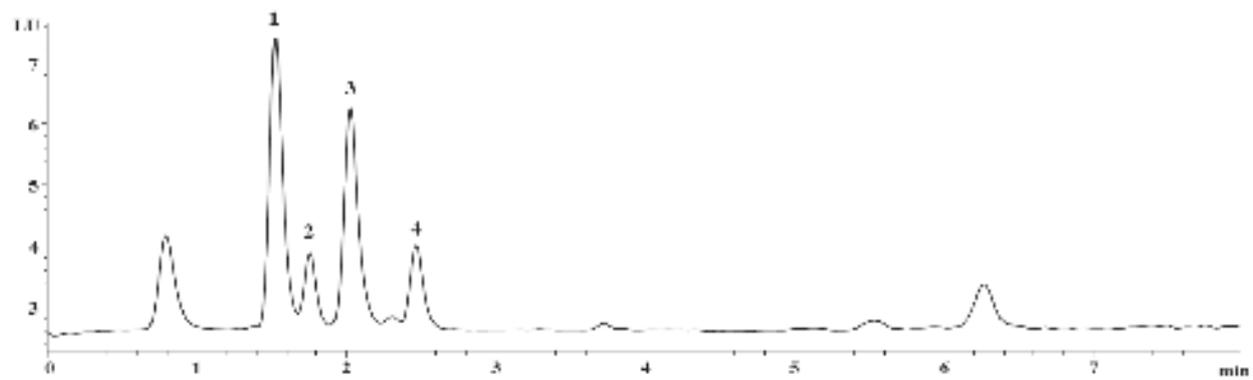

Figure 2.

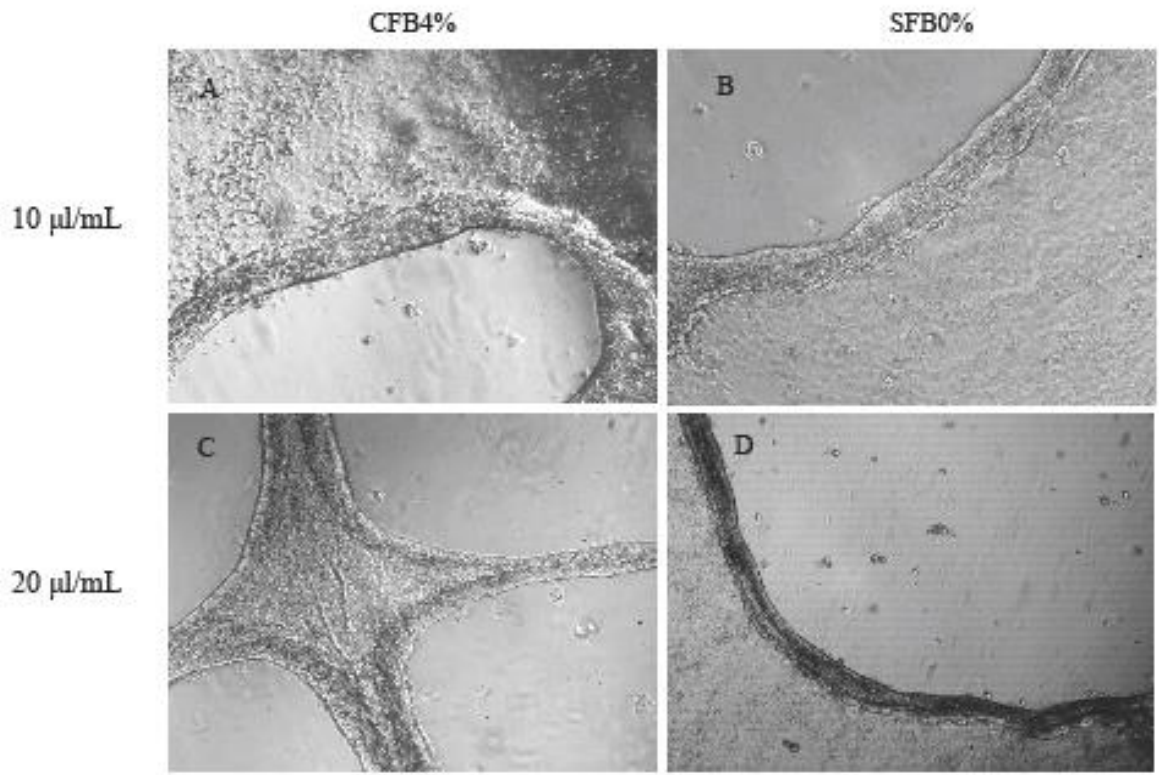

Figure 3.

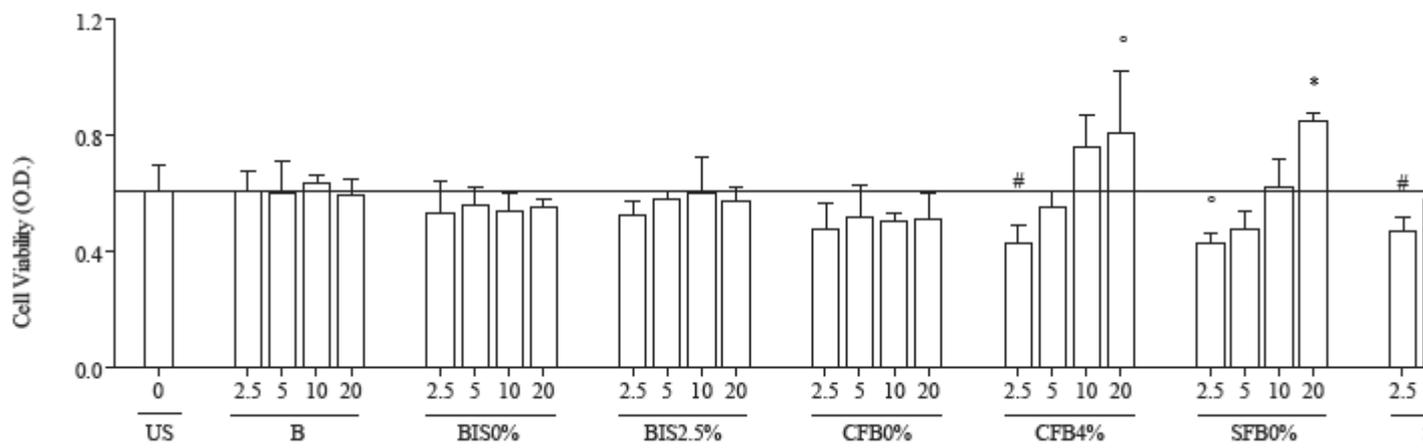

Figure 4.

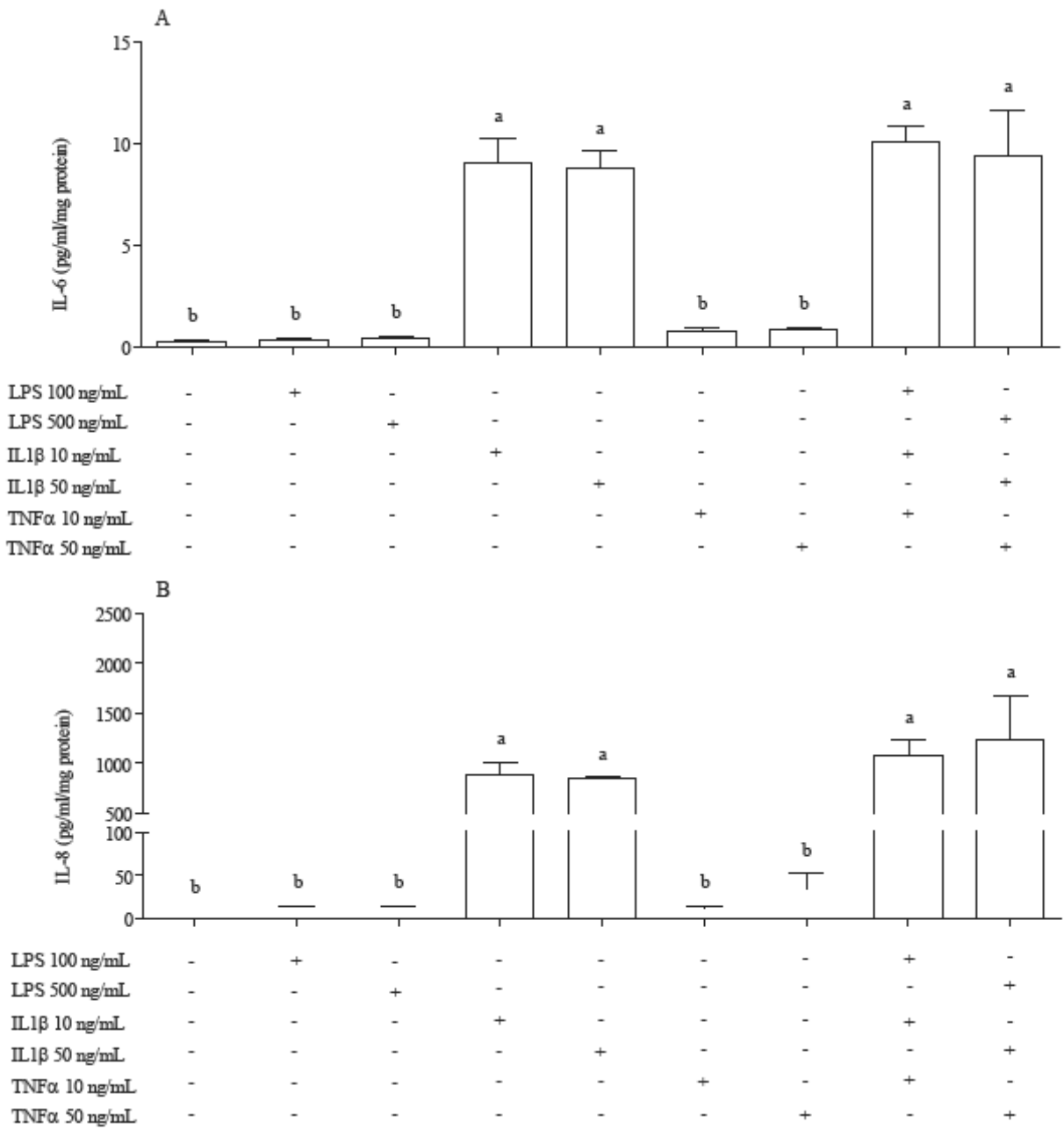

Figure 5.

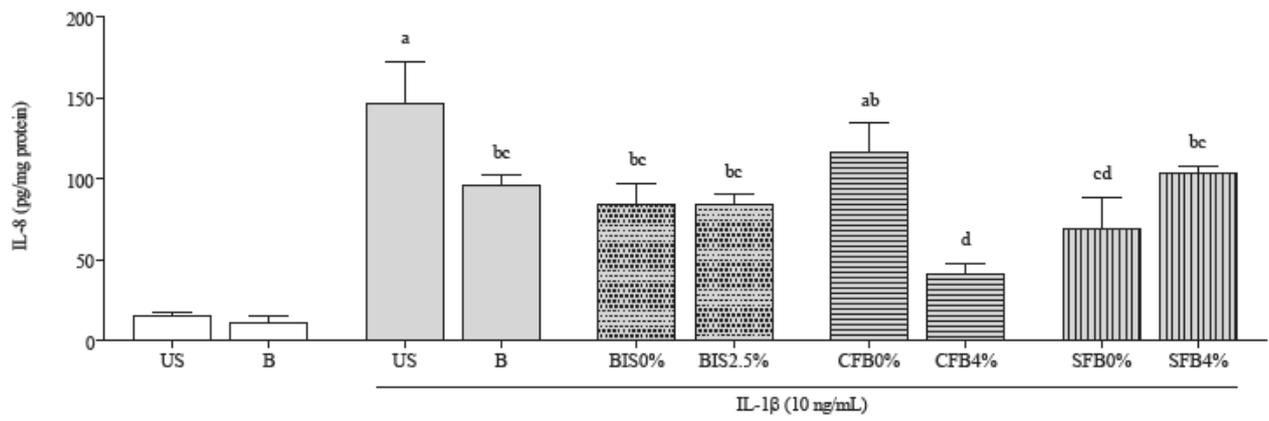

Figure 6.

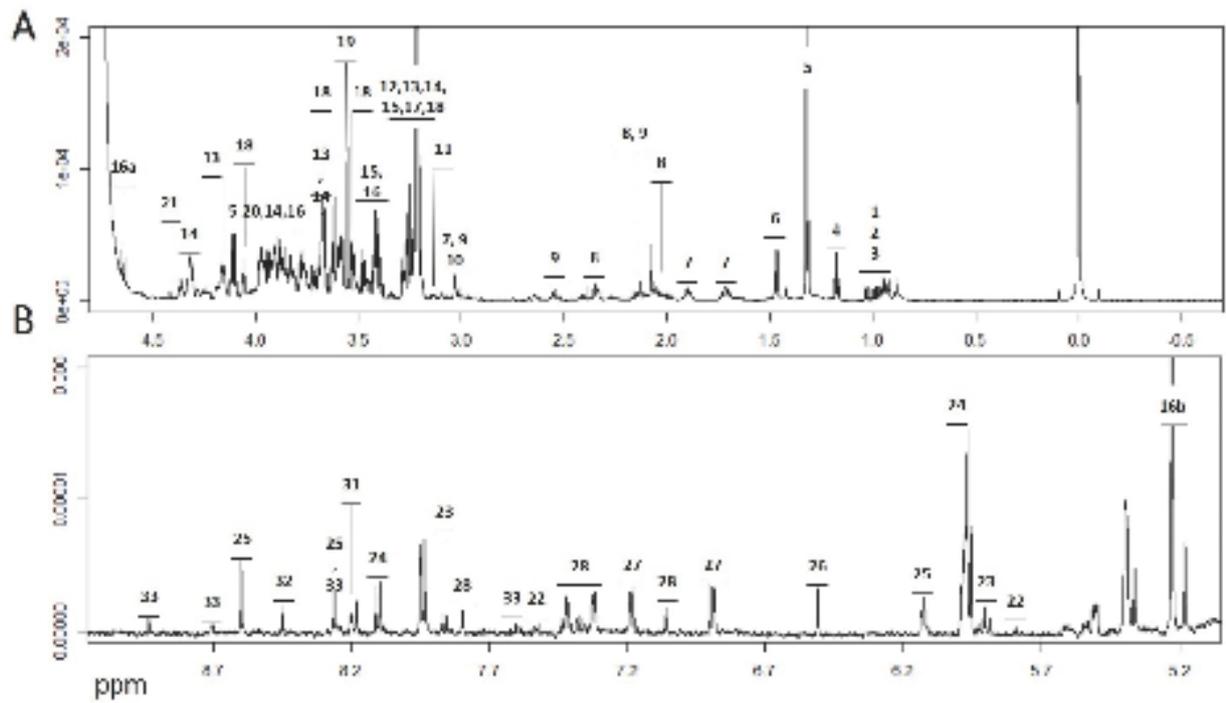

Figure 7.

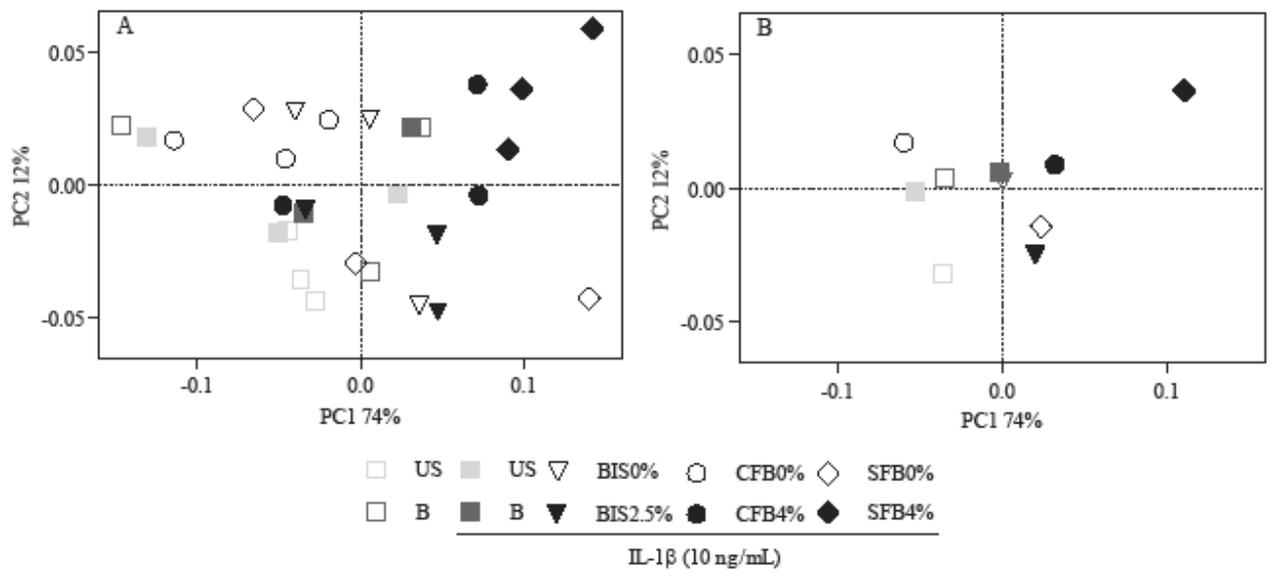

Figure 8.